\documentclass[twocolumn,prl,a4,aps]{revtex4}

\usepackage{graphicx}
\usepackage{dcolumn}
\usepackage{bm}

\newcommand{\qbar}{\overline{q}}
\newcommand{\rhoA}{{\rho^{(A)}}}

\renewcommand{\d}{\mathrm{d}}
\newcommand\Tr{\mathrm{Tr}}
\newcommand\ket[1]{|#1\rangle}
\newcommand\bra[1]{\langle#1|}

\newcommand\ent{\mathcal{E}}
\renewcommand\i{\mathrm{i}}

\begin{document}

\title{Entanglement in general two-mode continuous-variable states: local approach and mapping to a two-qubit system}

\author{H.-C. Lin}\email{ho.lin@ucl.ac.uk}
\author{A.J. Fisher}\email{andrew.fisher@ucl.ac.uk}

\affiliation{UCL Department of Physics and Astronomy and London Centre for Nanotechnology,\\ 
University College London, Gower Street, London WC1E 6BT, U.K.}

\begin{abstract}
We present a new approach to the analysis of entanglement in smooth bipartite continuous-variable states.  One or both parties perform projective filterings via preliminary measurements to determine whether the system is located in some region of space; we study the entanglement remaining after filtering.  For small regions, a two-mode system can be approximated by a pair of qubits and its entanglement fully characterized, even for mixed states.  Our approach may be extended to any smooth bipartite pure state or two-mode mixed state, leading to natural definitions of concurrence and negativity densities.  For Gaussian states both these quantities are constant throughout configuration space.
\end{abstract}


\maketitle

There has been growing interest in the quantification of entanglement in quantum systems.  However, many systems of interest have continuous, rather than discrete, degrees of freedom \cite{Braunstein.RevModPhys.77.531}.  The general characterization of entanglement in such continous-variable systems is a very difficult problem; most of the known results are for Gaussian states, where the logarithmic negativity \cite{PhysRevA.65.032314} can be calculated for arbitrary bipartite divisions  \cite{PhysRevA.66.042327}.  The entanglement of formation \cite{PhysRevA.54.3824} is known exactly only for symmetric bipartite Gaussian states \cite{giedke:107901}, and further measures of entanglement in multipartite Gaussian states are a topic of active current research \cite{Braunstein.RevModPhys.77.531,PlenioQuantPh0504163}.  For non-Gaussian states there is some progress in finding criteria for entanglement \cite{shchukin05}, but much less in quantifying it.

In this letter we present a new approach to this problem, allowing entanglement to be quantified locally in arbitrary (including non-Gaussian) continuous-variable states.  We concentrate on entanglement localized near particular regions in configuration space, which we analyse via a thought experiment in which the entangled state is first measured to localise it.  This corresponds to a particular type of projective filtering, used to identify the distribution of entanglement in a state which has a pre-existing bipartite structure.  This contrasts with previous studies \cite{botero04,plenio:060503,quantph0607069} of the entanglement of a finite region of space with the rest of the system.   We show that in the limit where the sizes of the measured regions become very small the description of each mode in the system becomes isomorphic to a single qubit.  In two-mode systems simple expressions result for entanglement monotones (concurrence and negativity), yielding natural definitions for corresponding densities in configuration space.

\textit{The filtering process.}  Let Alice and Bob share a state of two distinguishable one-dimensional particles (or two modes).  Alice can measure only the position of her particle or mode (coordinate $q_A$), Bob the position of his (coordinate $q_B$).  They filter their state by determining whether or not the particles are found in particular regions of configuration space, and discard instances in which they are not. We refer to the resulting sub-ensemble as the `discarding ensemble'.  For example, if Alice measures whether her coordinate is in a sub-region $a$, the corresponding projector is
\begin{equation}
\hat{E}_a=\int_a\ket{q_A}\bra{q_A}\,\d q_A\otimes \hat{1}_\mathrm{Bob}
\end{equation} 
where $\hat{1}_\mathrm{Bob}$ is the identity operation for Bob's particle. The density matrix $\hat{\rho}_D$ in the discarding ensemble after the measurement is
\begin{equation}\label{eq:definerhod}
\hat{\rho}_D(a)={\hat{E}_a\hat{\rho}\hat{E}_a\over p_a},
\end{equation}
where $p_a=\mathrm{Tr}[\hat{E}_{a}\hat{\rho}]$ is the probability
of finding Alice's particle in $a$.  Let the entanglement of $\hat{\rho}_D$ be $\ent_D$.

On the other hand if Alice chooses \emph{not} to discard the system when it is not in region $a$ (the `non-discarding ensemble'), the appropriate density operator $\hat{\rho}_{ND}$ is \begin{equation}\label{eq:rhond}
\hat{\rho}_{ND}=\hat{E}_{a}\hat{\rho}\hat{E}_{a}+\hat{E}_{a'}\hat{\rho}\hat{E}_{a'},\end{equation}
 where the complementary projector $\hat{E}_{a'}$ is defined as 
 \begin{equation}
\hat{E}_{a'}\equiv\hat{1}-\hat{E}_{a}=\int_{q_{A}\notin a}\d q_{A}\,\left|q_{A}\right\rangle \left\langle q_{A}\right|\otimes\hat{1}_{\mathrm{Bob}}.
\end{equation}
 Eq. (\ref{eq:rhond}) describes a mixed state, differing from the original $\hat{\rho}$
in that off-diagonal elements connecting $q_{A}\in a$
and $q_{A}\notin a$ have been set to zero.
If all the operators available to Alice have support only in region
$a$ (i.e. if she can neither measure her particle's properties, nor
manipulate it in any way, except when it is in $a$) then the second
component $\hat{E}_{a'}\hat{\rho}\hat{E}_{a'}$ contains no usable entanglement.  Alice can distinguish the two portions of $\hat{\rho}_{ND}$ by a local measurement, so the entanglement $\ent_{ND}$ in the non-discarding ensemble is just 
$\ent_{ND}=p_a\ent_{D}$.  We shall focus on calculating $\ent_D$ in this paper, noting that $\ent_{ND}$ can be simply obtained from it.  

\textit{Pure states: preliminary measurement on Alice's particle only.}
Suppose the state $\hat{\rho}$ is pure; so is $\hat{\rho}_D$.  It is therefore straightforward to calculate its entanglement from the von Neumann entropy of the corresponding reduced density matrix $\rhoA_D=\Tr_B[\hat{\rho}_D]$.  Suppose further that the initial filtering is performed only by Alice, by determining whether $q_A$ lies in the region $\qbar_A-a\le q_A\le\qbar_A+a$, and all instances in which this is not the case are discarded.  Now, since $a$ is to be very small, Alice's original reduced density matrix $\rhoA$ (before the measurement) in the neighbourhood of $\qbar_A$ can be expanded (provided it is smooth in configuration space) as  
\begin{eqnarray}\label{eq:rhoaexpand}
&&\rhoA(q_A,q'_A)=\rhoA(\qbar_A+x,\qbar_A+y)\\
&&\quad=\rhoA_{00}+\rhoA_{10}x+\rhoA_{01}y+\rhoA_{11}xy+\mathrm{O}(x^2,y^2),\nonumber
\end{eqnarray}
where
\begin{equation}
\rhoA_{nm}={\partial^n\over\partial {q_A}^n}{\partial^m\over\partial {q'_A}^m}\rhoA(q_A,q'_A)\big\vert_{q_A=q'_A=\qbar_A}.
\end{equation}
Within region $a$, $\rhoA_D$ is obtained by rescaling $\rhoA$ according to  (\ref{eq:definerhod}), where
$\Tr[\hat{E}_a\hat{\rho}]=2a[\rhoA_{00}+\mathrm{O}(a^2)]$.

Now seek right eigenfunctions $\phi_n$ of $\rhoA_D$ within the region $a$.
Expanding $\phi_n$ as a power series
\begin{equation}\label{eq:psipowerexpand}
\phi_n(x)=a_n+b_n x+\mathrm{O}(x^2),
\end{equation}
the eigenfunction condition becomes a matrix-vector equation operating on the expansion coefficients $\{a,b,\ldots\}$. 
To order $a^2$, the non-zero eigenvalues are:
\begin{eqnarray}
\lambda_1&=&1-\lambda_2\nonumber\\
\lambda_2&=&{a^2\over 3\rhoA_{00}^2}(\rhoA_{11}\rhoA_{00}-\rhoA_{01}\rhoA_{10}).
\end{eqnarray}
So to lowest order ($a^2$) the eigenvalues, and hence the von Neumann entropy, of $\rhoA_D$ are entirely determined by the quantity $\epsilon\equiv\lambda_2$;
the corrections due to higher eigenvalues, or to higher-order terms in Eq.~(\ref{eq:rhoaexpand}), affect the result only to order $a^4$.
Specifically, the von Neumann entropy is
\begin{equation}
S=h(\epsilon)\equiv-[\epsilon\log_2(\epsilon)+(1-\epsilon)\log_2(1-\epsilon)].
\end{equation}
Note that if Alice's state is pure, $\rhoA_{11}\rhoA_{00}=\rhoA_{01}\rhoA_{10}$,
so $S$ is zero as we would expect.  

\textit{Pure states: preliminary measurement on both particles.}
It is possible to generalise this analysis to the case where both Alice and Bob make a prelminary measurement to localise their particles, within regions $\{q_A:\qbar_A-a\le q_A\le\qbar_A+a\}$ and $\{q_B:\qbar_B-b\le q_B\le\qbar_B+b\}$ respectively.  In that case one can expand $\rho$ as a joint power series in $\{q_A,q'_A,q_B,q'_B\}$, calculate the reduced density matrix $\rhoA$ (also as a power-series expansion) and proceed as above.  However, further insight can be obtained by an alternative approach.  Define for both Alice and Bob two-dimensional state spaces consisting of 
\begin{eqnarray}\label{eq:alicebobbasis}
\phi_{A0}(x_1)=\sqrt{1\over 2a};\qquad \phi_{A1}(x_1)=\sqrt{3\over 2a^3}x_1;\nonumber\\
\phi_{B0}(x_2)=\sqrt{1\over 2b};\qquad \phi_{B1}(x_2)=\sqrt{3\over 2b^3}x_2,
\end{eqnarray}
which are orthonormal on the intervals $-a<x_1<a$ and $-b<x_2<b$ respectively; $\phi_0$ represents the constant component of the wave function, and $\phi_1$ the spatially varying part.  So long as terms varying as $x^2$ or higher can be neglected, a Taylor expansion of the joint state to linear order (\ref{eq:rhoaexpand}) is equivalent to expanding $\psi$ in the basis (\ref{eq:alicebobbasis}), thereby reducing the joint system to a two-qubit one.  It can be shown that the third largest eigenvalue of $\rhoA$ (corresponding to the extent to which the two-qubit approximation fails) is now of order $(ab)^4$.

We can now use any of the standard measures of the entanglement of the two-qubit system.  For pure states,  the tangle \cite{wootters98b} is
\begin{equation}\label{eq:tangle}
\tau=\frac{1}{4}\left({1\over ab|\psi_{00}|^2}\right)^2\left|{4a^2b^2\over 3}(\psi_{00}\psi_{11}-\psi_{01}\psi_{10})\right|^2,
\end{equation}
where 
\begin{equation}
\psi_{nm}\equiv{\partial^n\over\partial {q_A}^n}{\partial^m\over\partial {q_B}^m}\psi(q_A,q_B)\big\vert_{q_A=\qbar_A,q_B=\qbar_B}.
\end{equation}
The prefactor in (\ref{eq:tangle}) comes from the normalization condition $\int_{-a}^a\d x_1\int_{-b}^b\d x_2|\psi(x_1,x_2)|^2=1$.
The entanglement is therefore $h\left((1-\sqrt{1-\tau})/2\right)=h\left(\tau/4+\mathrm{O}(\tau^2)\right)$.  By analogy with the definition of concurrence $\cal{C}=\sqrt{\tau}$ for two-qubit states \cite{wootters98}, we define a \textit{concurrence density} $c\equiv{\cal C}/(ab)$ such that $\tau=(cab)^2$; then
\begin{eqnarray}\label{eq:concurrencedensity}
c&=&{2\over3\rho_{0000}}[\rho_{1100}\rho_{0011}
+\rho_{0000}\rho_{1111}\nonumber\\
&&\quad-\rho_{1000}\rho_{0111}-\rho_{0100}\rho_{1011}]^{1/2},
\end{eqnarray}
where
\begin{eqnarray}
\rho_{ijkl}&\equiv&{\partial^i\over\partial {q_A}^i}{\partial^j\over\partial {q'_A}^j}{\partial^k\over\partial {q_B}^k}{\partial^l\over\partial {q'_B}^l}\rho(q_A,q_B;q'_A,q'_B)\big\vert_{\qbar_A,\qbar_B}\nonumber\\
&=&\psi_{ik}\psi^*_{jl}
\end{eqnarray}
for pure states.

The negativity $\mathcal{N}$ of the filtered state can also be computed; we define this as the sum of the magnitudes of the negative eigenvalues $\lambda_i$ of the partially transposed density matrix $\rho^{T_B}$,
\begin{equation}
\mathcal{N}=\sum_{i\,\textbf{s.t.}\,\lambda_i<0}|\lambda_i|.
\end{equation}
For pure states negatitivity and concurrence are simply related \cite{JModOptic46.145,PRL95.040504}: $\mathcal{N}={\cal C}/2$.  

The accuracy of the two-qubit approximation is guaranteed (for sufficiently small $a$ and $b$) by the fact that each party's reduced density matrix has only two non-zero eigenvalues of the density matrix to order $(ab)^2$.  

\textit{Mixed states.} The mapping to a two-qubit system applies also to mixed states, where exact recipes for the entanglement of formation \cite{wootters98} and other measures are known.  We find all eigenvalues $\mu_i$ of $\rho\tilde{\rho}$ (as defined in \cite{wootters98}) are at leading order proportional to $(ab)^2$ so again there is a well-defined concurrence density $c$. These leading terms (and hence the concurrence ${\cal C}$) can be found by solving a quartic, although its roots are not simple in general.

Particularly simple expressions can be found for the local negativity of a mixed state.  We define the quantities
\begin{eqnarray}
D_{1} & = & (\rho_{1100}\rho_{0000}-\rho_{0100}\rho_{1000})/(3\rho_{0000}^{2});\nonumber\\
D_{2} & = & (\rho_{0011}\rho_{0000}-\rho_{0001}\rho_{0010})/(3\rho_{0000}^{2});\nonumber\\
C_{1} & = & \left[(\rho_{0000} \rho_{0101}-\rho_{0001} \rho_{0100}) \rho_{1010}\right.\\
   &&\quad+\rho_{0011} (\rho_{0100} \rho_{1000}-\rho_{0000}
   \rho_{1100})\nonumber\\
   &&\quad\left.+\rho_{0010} (\rho_{0001} \rho_{1100}-\rho_{0101} \rho_{1000})\right]/(9\rho_{0000}^{3}).\nonumber
\end{eqnarray}
If $C_1\ge 0$, the negativity depends on $a$ and $b$ individually, not just on the product $ab$.  If we seek the maximum negativity while keeping $ab$ fixed, we find this occurs when $a^{2}D_{1}=b^{2}D_{2}$, and obtain a corresponding \textit{negativity density} $n=\mathcal{N_\mathrm{max}}/(ab)$ given by
\begin{equation}\label{eq:negativitydensity} 
n=\sqrt{C_{1}+D_{1}D_{2}}-\sqrt{D_{1}D_{2}}.
\end{equation}
For pure states, $D_{1}$ and $D_{2}$ vanish, and we recover the pure-state negativity expected from equation~(\ref{eq:concurrencedensity}).  If $C_{1}<0$ there is no negativity to order $(ab)$; the full treatment of the negativity will be discussed in a subsequent paper.

\textit{Example: Entanglement of Gaussian states.}  
The \textit{characteristic
function} $C$ is defined in terms of the Weyl operator $\hat{C}$ (taking $\hbar=1$) through
\begin{equation}
C(X,P)=\mathrm{Tr}(\hat{\rho}\hat{C}(X,P));\quad\hat{C}(X,P)=e^{\mathrm{i}(X\hat{q}+P\hat{p})},
\end{equation}
where the position operator is denoted by $\hat{q}$
and the momentum operator by $\hat{p}$.  A state $\hat{\rho}$ is said to be Gaussian when its characteristic function
is a Gaussian in phase space.  This important set of states includes both thermal and `squeezed' states of harmonic systems and plays a key role in several fields of theoretical and experimental
physics; we use them as an example because their entanglement properties are better understood than those of other continuous-variable systems, while recognising that our approach is general.  The corresponding configuration-space density matrix can be written \cite{PhysRevA.36.3868} 
\begin{eqnarray}\label{eq:generalgaussian}
&&\rho(q;q')=\zeta_{1}\mathrm{exp}\bigl[-q^{T}\mathbf{L}q-q'^{T}\mathbf{L}q'\\
&&\quad-\frac{1}{2}(q-q')^{T}\mathbf{M}(q-q')
+\frac{i}{2}(q-q')^{T}\mathbf{K}(q+q')\bigr],\nonumber\end{eqnarray}
where $\zeta_{1}$ is a normalization constant, and where $\mathbf{L}$, $\mathbf{M}$
and $\mathbf{K}$ are real $N\times N$ matrices (for $N$ modes) with $\mathbf{L}$ and $\mathbf{M}$ symmetric,
while $\mathbf{K}$ is arbitrary.  
For thermal states of two similar but distinguishable oscillators each having mass $m$, angular frequency $\omega$ and coupling spring constant $K$ with corresponding dimensionless coupling
$\alpha={2K}/m\omega^{2}$, $\mathbf{K}=0$ and the values of $\mathbf{L}$ and $\mathbf{M}$ are given by \cite{PhysRevA.36.3868,PhysRevA.66.042327}.  We adopt $(m\omega)^{-1/2}$ as our length unit.

First we consider the ground state.  Eq. (\ref{eq:concurrencedensity}) becomes
\begin{equation}\label{eq:guassiandensity}
c=\frac{\sqrt{2}m\omega}{6}\sqrt{1+2\alpha-\sqrt{1+4\alpha}}.
\label{eq:Gaussian-2}\end{equation}
So, for a given small $a$ and $b$, the entanglement depends only 
on $\alpha$ and the fundamental length unit; it is \textit{independent} of the location of the centres $(\qbar_A,\qbar_B)$ of the measurement regions.  The lack of dependence on $(\qbar_A,\qbar_B)$ is a special feature of Gaussian states, and this result does not hold in general; note also that the concurrence density can be made arbitrarily large by increasing $\alpha$.

\begin{figure}[t]
\begin{tabular}{cc}
\includegraphics[width=4cm]{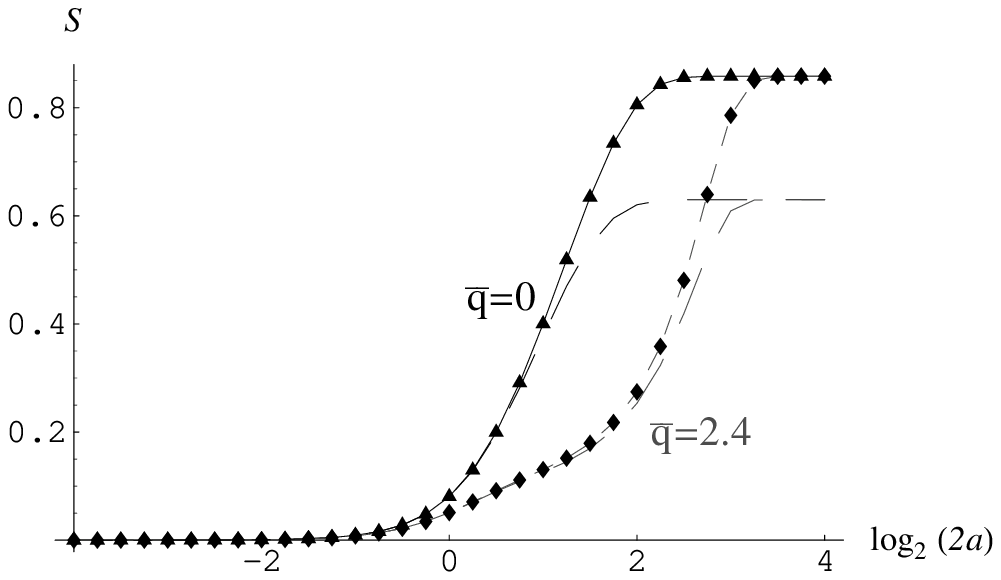}&
\includegraphics[width=4cm]{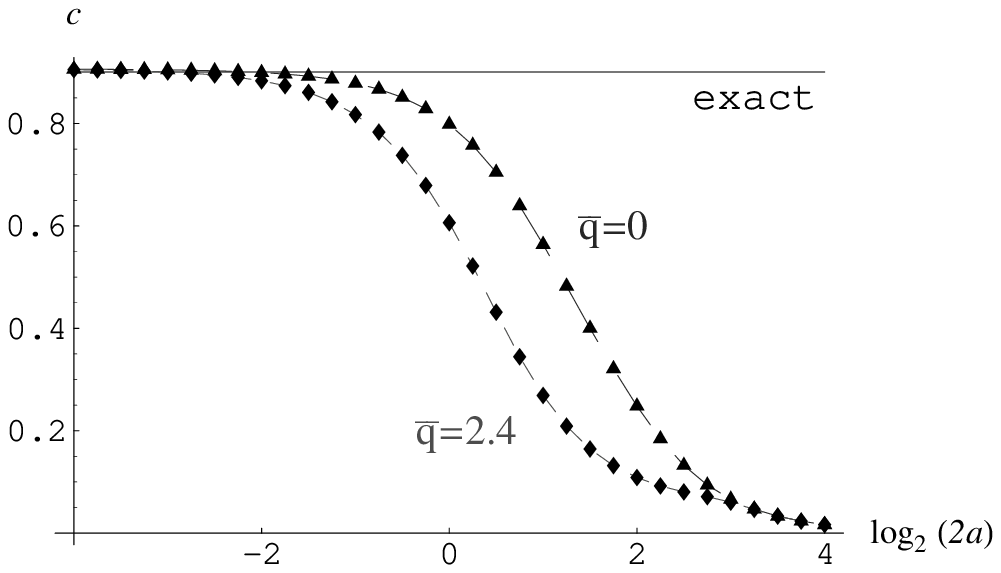}\\
(a)&(b)\\
\end{tabular}
\caption{\label{fig:concurrencevsregionsize}
Entanglement properties as a function of region size for a Gaussian ground state with $\alpha=10$ and $m=\omega=1$, in the case where both Alice and Bob make preliminary measurements and the reigion sizes are chosen to be the same.  (a) Entanglement $S$ (dimensionless)  as a function of region size $2a$ (in units of $(m\omega)^{-1/2}$; log scale) for two different positions (data points); the entanglements contained in the effective two-level systems constructed from the two largest eigenvalues of $\rhoA$ are also shown (dashed lines).  (b) The concurrence density $c$ (in units of $m\omega$), computed from the entanglement of formation by inverting the relation $S=h((cab)^2/4)$; note how it saturates to the exact result predicted by (\ref{eq:guassiandensity}) (horizontal line) for small regions.}
\end{figure}

Figure~\ref{fig:concurrencevsregionsize} shows the variation of the entanglement $S$ (calculated numerically) with the region size.  Note how the entanglement saturates to the full value given in \cite{rendell:012330} for large regions, while for small regions it reduces to the value predicted by (\ref{eq:guassiandensity}).  To obtain entanglements of a substantial fraction of one ebit, it is necessary to choose a region size comparable to the fundamental length unit of the oscillator; around  this point the two-qubit approximation is just starting to break down.  Calculational details and further results are described in \cite{quantph0609078}.
 
\begin{figure}[t]

\includegraphics[width=6cm]{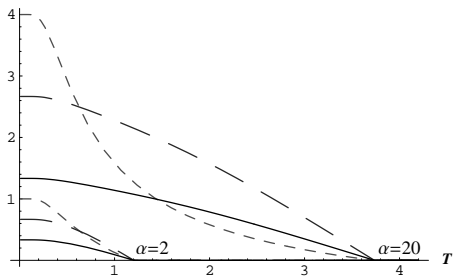}

\caption{\label{fig:guassianresults}
Negativity density $n$ (full lines; in units of $m\omega$ ) and concurrence density $c$ (long dashes; in units of $m\omega$) as a function of temperature $T$ (in units of $\omega$)  for a thermal state of the two-oscillator system discussed in the text having $k_B=m=\omega=1$, and for two different values of the coupling $\alpha$.  The global negativity $\mathcal{N}_{g}$ \cite{PhysRevA.66.042327} is shown for comparison (short dashed lines; dimensionless).}
\end{figure}

For mixed Gaussian states, we find that the concurrence density and negativity density are again independent of position.  In Figure~\ref{fig:guassianresults} we plot both quantities as a function of temperature for thermal states of two oscillators with different coupling strengths.  We also show the conventional (global) negativity for this state; note that both local entanglement densities vanish at the same temperature as the global negativity, showing that for this set of  states, those which are entangled from the global point of view are also entangled by our local measures.  It would be very interesting to know whether this property applies more generally.  We also note that the concurrence density is exactly twice the negativity density; this is a special property of two-mode Gaussian states, which will be discussed further elsewhere.

\textit{Extraction of the local entanglement.}  Methods of extracting the entanglement from a squeezed continuous-variable state into a pair of two-level system were previously studied in \cite{JModOptic49.1739,PRA70.022320,PRL92.013602}  Using our mapping to an effective two-qubit system, we can swap the entanglement in a small region of the continuous wavefunction to local qubits (i.e.\ true two-level systems). Remembering that the states $\phi_0$ and $\phi_1$ drop to zero outside the region $[-a,a]$ we find that the Pauli operators $\hat{X}$ and $\hat{Y}$ of the effective qubit can be represented in terms of the canonical position and momentum operators $\hat{q}$ and $\hat{p}$ by
\begin{equation}
\hat{X}={\sqrt{3}\over a}\hat{q};\qquad \hat{Y}=-{2 a\over\sqrt{3}\hbar}\hat{p}.
\end{equation}
The experiment could be performed as follows: first, localize the continuous degree of freedom (for example, through a homodyne measurement in the case of an electromagnetic field mode), then perform a SWAP operation by composing three controlled-X gates:
\begin{eqnarray}
\hat{U}_{SWAP}&=&\exp\left[{\i\pi\over 4}(\hat{\sigma}_y-1)({\sqrt{3}\over a}\hat{q}-1)\right]\nonumber\\
&&\times\exp\left[{\i\pi\over 4}(\hat{\sigma}_x-1)(-{2 a\over\sqrt{3}\hbar}\hat{p}-1)\right]\nonumber\\
&&\times\exp\left[{\i\pi\over 4}(\hat{\sigma}_y-1)({\sqrt{3}\over a}\hat{q}-1)\right],
\end{eqnarray}
where $(\hat{\sigma}_x,\hat{\sigma}_y)$ are Pauli operators for a local qubit.  
Using this procedure one could therefore extract the full two-level entanglement shown in the dashed curves of Figure~\ref{fig:concurrencevsregionsize}(a) provided the swap operation is successful.

\textit{Discussion.}  After filtering in configuration space as we have described, it is simple to characterize the entanglement in a continuous-variable system.  One could generalize our results to multimode systems by making multivariate Taylor expansions of $\hat{\rho}$ or by using the isomorphism of each mode to one qubit.  Our approach leads naturally to quantities characterizing the entanglement that scale with the extent of the measurement region used in the filtering, and hence to the identification of well-defined densities.  However, the concurrence and negativity are not extensive, in the sense that the sum of these quantities over all the sub-regions of configuration space does not yield the full entanglement of the original system.  This point is discussed further in \cite{quantph0609078}.  It will be interesting to characterize further the relationship between the local and global views of continuous-variable entanglement; in any case our results open a wide range of non-Gaussian states to further study.

\textit{Acknowledgments.}  We are grateful to Sougato Bose and Alessio Serafini for a number of valuable discussions.

\bibliographystyle{apsrev}
\bibliography{entanglement_density_HCL_AJF_reference}

\end{document}